\documentclass[usenatbib]{mn2e}

\usepackage{amsmath}
\usepackage{natbib}
\usepackage{epsfig}
\usepackage{txfonts}

\newcommand{\beq}{\begin{equation}}   %

\newcommand{\eeq}{\end{equation}}   %

\newcommand{\beqa}{\begin{eqnarray}}   %

\newcommand{\eeqa}{\end{eqnarray}}   %

\newcommand{\beal}{\begin{align}}

\newcommand{\enal}{\end{align}}

\newcommand{\bspl}{\begin{split}}

\newcommand{\espl}{\end{split}}

\newcommand{\bsub}{\begin{subequations}}

\newcommand{\esub}{\end{subequations}}

\newcommand{\bmulti}{\begin{multline}}   %

\newcommand{\beqm}{\begin{mathletters}}   %

\newcommand{\eeqm}{\end{mathletters}}   %

\usepackage{color}

\usepackage{hyperref}
\usepackage{grffile}
\usepackage{graphics}
\usepackage{booktabs}

\title[Spatial variation of $r$]
{Linking the BICEP2 result and the hemispherical power asymmetry through spatial variation of $r$}

\author[Chluba et al.]{
 J.~Chluba$^{1}$\thanks{E-mail: jchluba@pha.jhu.edu}, Liang Dai$^{1}$, Donghui Jeong$^{1, 2, 3}$,
 Marc Kamionkowski$^1$ and Amanda Yoho$^{4}$  
 \\
$^{1}$ Department of Physics and Astronomy, Johns Hopkins University, Bloomberg Center, 
3400 N. Charles St., Baltimore, MD 21218, USA
 \\
$^{2}$ Department of Astronomy and Astrophysics, The Pennsylvania State University, University Park, PA 16802
\\
$^{3}$ Institute for Gravitation and the Cosmos, The Pennsylvania State University, University Park, PA 16802
 \\
$^{4}$ Department of Physics, Case Western Reserve University,  
2076 Adelbert Road, Cleveland, Ohio 44106-7079, USA
}
\date{{\vspace{-2mm}
Accepted 2014 May 6. Received 2014 April 10.}}

\begin{document}

\maketitle

\begin{abstract}
For the simplest inflation models, the recent detection of a large primordial B-mode polarization signal by the BICEP2 experiment indicates a slight tension with the upper limit on the tensor-to-scalar ratio, $r$, from the Planck satellite. Here, we discuss spatially varying $r$ as a possible explanation for this discrepancy. This idea seems attractive since it may also explain part of the hemispherical temperature power asymmetry seen by WMAP and Planck at large angular scales. If these two aspects are indeed connected, the model suggests that in the Northern hemisphere $r$ should be much smaller, a hypothesis that could be confirmed with future B-mode experiments, providing a test for the stationarity of primordial tensor contributions across the sky. The BICEP2 measurement furthermore rules out that a simple dipolar modulation of $r$ alone can be responsible for the full hemispherical power asymmetry.
\end{abstract}


\begin{keywords}
Cosmology: cosmic microwave background -- theory -- observations
\end{keywords}

\section{Introduction}
\label{sec:intro}
The detection of large primordial B-modes by the BICEP2 experiment, with a tensor-to-scalar ratio $r=0.2^{+0.07}_{-0.05}$  \citep{BICEP2results}, has excited the cosmology community for the past month. 
Not only does this large value for $r$ suggest that sub-orbital B-mode experiments like SPIDER \citep{SPIDER}, CLASS \citep{CLASS}, Polarbear \citep{Polarbear}, SPTpol \citep{SPTpol} and ACTpol \citep{ACTPol} should be able to characterize the B-mode power spectrum to high precision over the next few years, but it also points towards a slight tension with the upper limit $r<0.11$ (95\% c.l.) deduced from measurements of the temperature power spectrum by the Planck team \citep{Planck2013params}. One simple extension that restores the consistency between these measurements is to allow for a small negative running of the scalar spectrum, pushing us into the regime of non-standard inflation scenarios, since the simplest slow-roll models predict negligible running. 

These recent findings have spurred much discussion. Could the large B-modes be due to foregrounds or some unaccounted temperature-polarization leakage \citep{Liu2014}? Is the tensor power spectrum blue-tilted \citep{Brandenberger2014, Gerbino2014, Biswas2014}? Do the BICEP2 results require non-standard inflation scenarios or more general early-universe models \citep{Harigaya2014, Nakayama2014, Contaldi2014, Abazajian2014, Miranda2014, McDonald2014, Scott2014}? Should we worry about large-field excursions \citep{Kehagias2014, Choudhury2014, Lyth2014} violating the Lyth bound? Maybe primordial magnetic fields rather than gravity waves generate the B-mode signal \citep{Bonvin2014}? Is a sterile neutrino the culprit \citep{Zhang2014, Dvorkin2014}? What about topological defects \citep{Lizarraga2014, Moss2014}? Clearly, more data are needed to refine the polarized foreground model and further tighten the constraints on the B-modes, answering these questions, and the community is eagerly awaiting the next round of results from Planck, SPTpol and the Keck array.

In this paper, we suggest yet another possible explanation for the large value of $r$ found by BICEP2 that is also consistent with the all-sky upper limit from Planck. The idea is motivated by the fact that the BICEP2 footprint is about $60$ degrees away from the maximum of the hemispherical power asymmetry \citep{Eriksen2004, Bennett:2010jb, Planck2013power, Aslanyan2013, Akrami2014}. As suggested by \citet{Dai2013}, a spatial variation of $r$ could provide one viable explanation for at least part of the temperature power asymmetry and its scale dependence \citep{Hirata2009power, Flender2013}. This could, e.g., be caused by an exotic super-horizon tensor mode \citep{Abolhasani2013}, a modulated preheating scenario \citep{Bethke2013}, dissipative processes \citep{DAmico2013}, or more generally in multi-field inflation models that possibly independently generate scalar and tensor perturbations. 

One expectation is that a detection of primordial B-modes will be easier in the Southern hemisphere since there the value for $r$ lies above the average. This also suggests that in the Northern hemisphere the tensor contribution should be much smaller. In the future, this hypothesis could be tested by future B-mode experiments with sufficient sky-coverage, providing a check for the stationarity of the tensor contribution across the sky, even probing cases beyond a simple dipolar power asymmetry.

\vspace{-3.5mm}

\section{Linking the power asymmetry to spatially varying tensor modes}
\label{sec:spatial_var_r}
The hemispherical asymmetry is consistent with a dipolar modulation of an otherwise statistically isotropic cosmic microwave background (CMB) sky \citep{Prunet2005, Gordon2005, Gordon2007}, where the best-fitting dipole in galactic coordinates has a direction $(l, b) \approx  (227, -27)^\circ$ and amplitude (in terms of r.m.s. temperature fluctuations on large angular scales, multipoles $\ell \lesssim 64$) of $A = 0.072 \pm 0.022$ \citep{Planck2013power}. 
Explicitly, the CMB temperature fluctuation in a direction $\hat n$ can be written as $\Delta T(\hat n)=\Delta T_{\rm iso}(\hat n)[1+A \,\hat n\cdot \hat p]$ in this case, where $\Delta T_{\rm iso}(\hat n)$ denotes the temperature fluctuation for a statistically isotropic sky and $\hat p$ defines the dipole axis. Then $A$ can be defined as $A \simeq (1/2)[\sum_{\ell=2}^{\ell_{\rm max}}(2\ell +1) \Delta C_\ell^{TT}/C_\ell^{TT}]/\sum_{\ell=2}^{\ell_{\rm max}}(2\ell +1)$, where $\Delta C_\ell^{TT}/C_\ell^{TT}$ is the fractional correction to the CMB temperature power spectrum with respect to the sky average \citep[see][for more details]{Dai2013}.

In galactic coordinates, the central region of the BICEP2 footprint lies at $(l, b) \simeq  (310, -59)^\circ$, which is roughly $60^\circ$ away from the power maximum. Assuming that $r$ varies spatially as $r(\theta)=r_0+\Delta r \cos\theta$, with $\theta$ defining the angle relative to the maximum of the hemispherical power asymmetry, this suggests $r_{\rm BICEP}\approx r_0+ \Delta r/2$. Assuming $r_0\simeq \Delta r\simeq 0.11$ one thus finds $r_{\rm BICEP}\approx0.17$, consistent with the BICEP2 result. This model by construction is also consistent with the Planck all-sky constraint. It furthermore suggest that in the direction $(l, b) \approx  (227, -27)^\circ$ the contribution of tensor modes is close to $r_{\rm max}\approx 0.22$, while in the opposite direction $r_{\rm min}\approx 0$, a hypothesis that can be checked by future B-mode experiments. For this it will be important to distribute the measurements in both hemispheres, sampling a sufficient fraction of the whole sky, and also by combining different experiments. In the near future, this question could potentially be addressed by CLASS and SPIDER, which independently cover large parts of the CMB sky. Looking farther ahead, a CMB polarization measurement from space using PIXIE \citep{Kogut2011PIXIE}, LiteBird \citep{LiteBIRD} or a mission similar to PRISM \citep{PRISM2013WPII} could also allow testing this scenario.

\begin{figure}
\centering
\includegraphics[width=0.94\columnwidth]{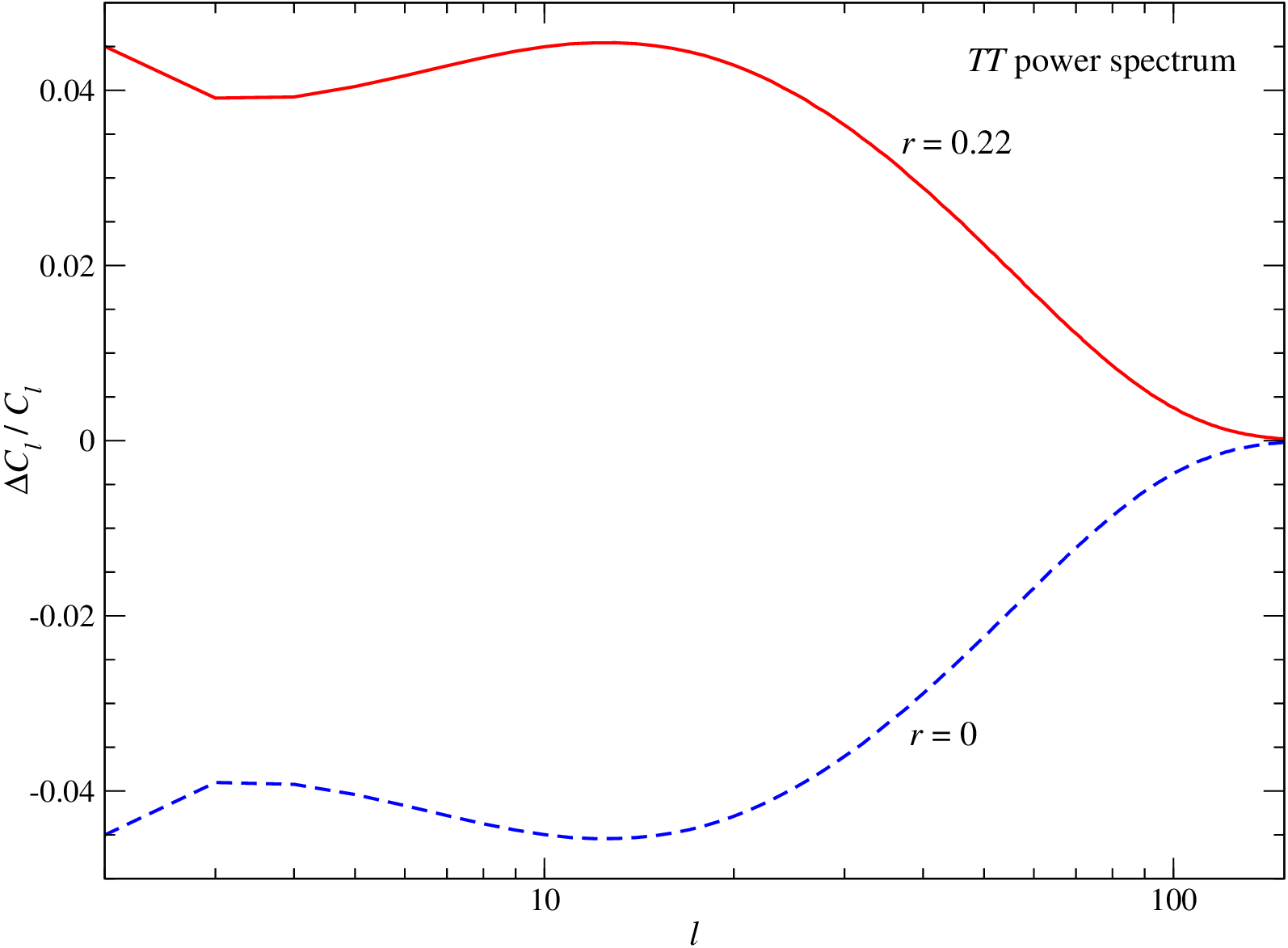}
\\[1.5mm]
\includegraphics[width=0.94\columnwidth]{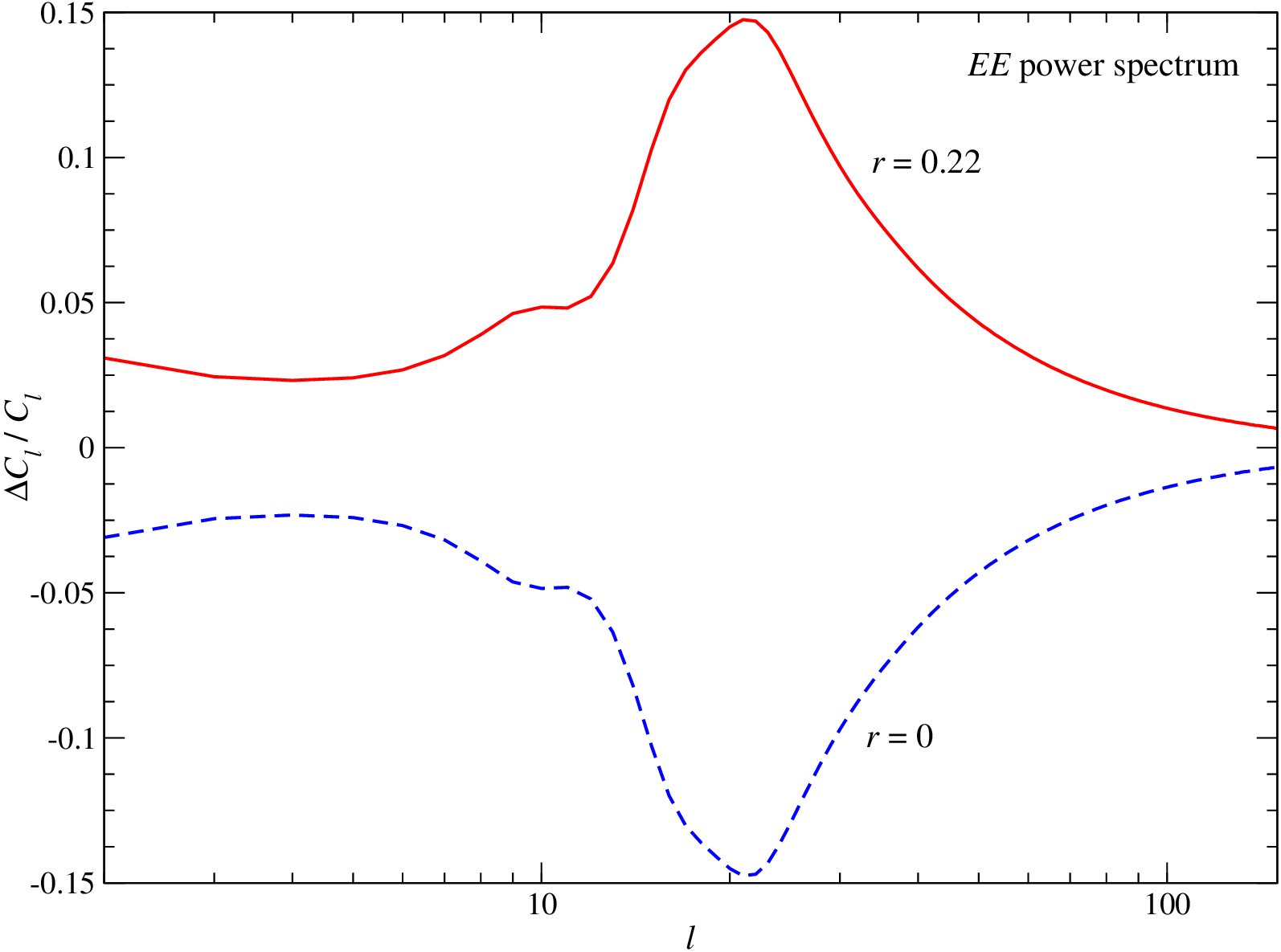}
\\[1.5mm]
\includegraphics[width=0.94\columnwidth]{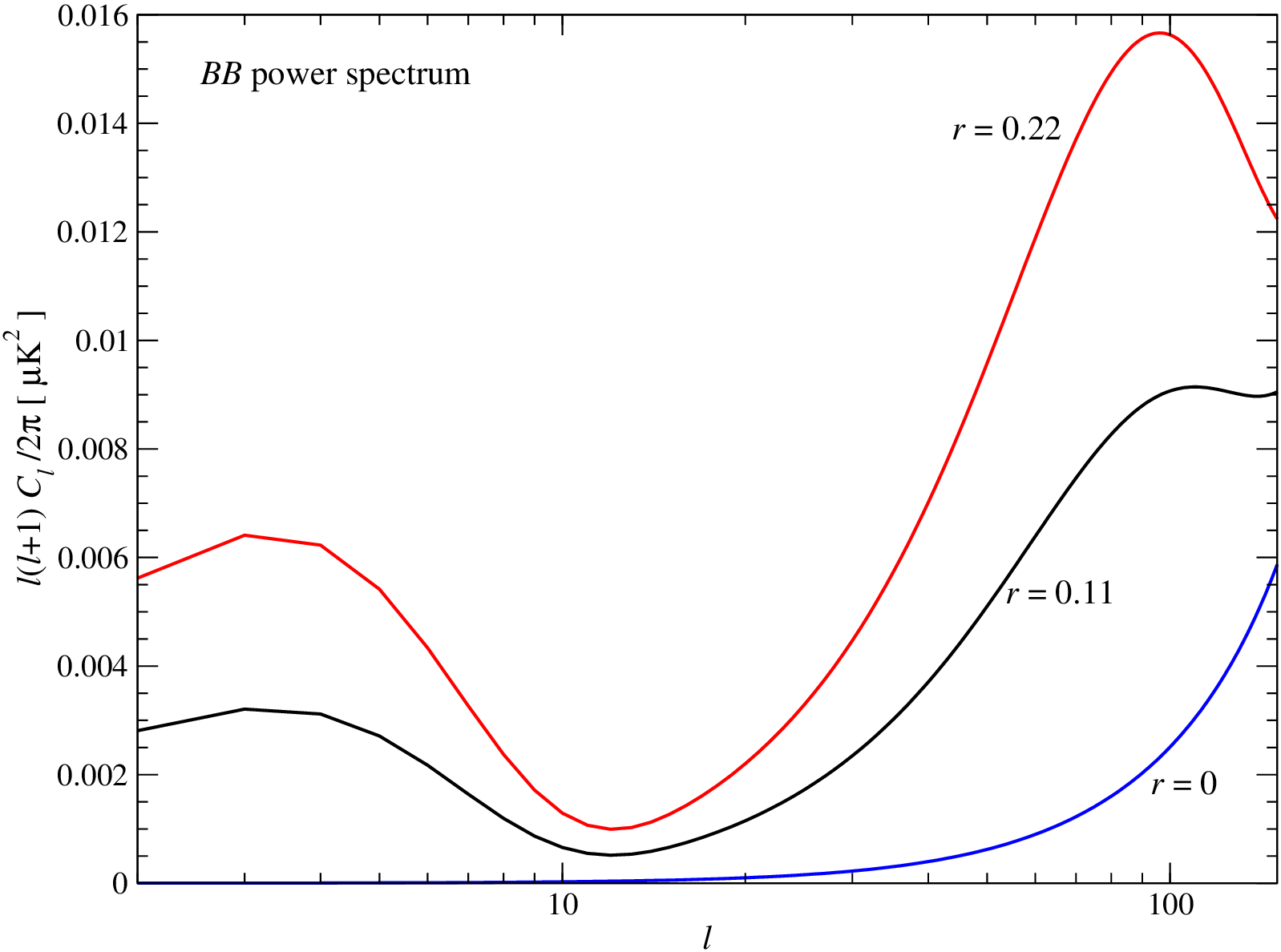}
\caption{Relative changes in the $TT$ and $EE$ power spectra caused by the spatial variation of $r$ (upper two panels) with respect to the all-sky average with $r\simeq 0.11$. The lower panel directly shows the $BB$ power spectrum for $r=0, 0.11$ and $0.22$. If a dipolar modulation of $r$ is present, measurements of the $EE$ and $BB$ power spectra will add additional direct information. The curves were obtained using CAMB \citep{CAMB} for the Planck cosmology \citep{Planck2013params}.}
\label{fig:CMB_TT}
\end{figure}
In Fig.~\ref{fig:CMB_TT}, we show the temperature and polarization power spectra at large angular scales ($\ell\lesssim100$). For the $TT$ and $EE$ power spectra the relative differences with respect to the sky average are shown, while for $BB$ we show the power spectrum directly.
Using $r_{\rm max}\approx 0.22$ and $r_{\rm min}\approx 0$, at $\ell\lesssim 64$ we find an overall power asymmetry amplitude of $A\simeq 0.016$, which explains part of the power asymmetry found by the Planck team \citep{Planck2013power}. To explain the full power asymmetry, one requires $r_0\simeq \Delta r \simeq 0.65$ \citep{Dai2013}, which is already in strong tension with the all-sky average from Planck. It would furthermore predict $r_{\rm BICEP}\simeq 0.94$, which is ruled out by the BICEP2 measurement at more than $10 \sigma$. We note that $r_0\simeq \Delta r$ maximizes the overall tensor modulation, while more generally $0<\Delta r< r_0$ gives models with $r>0$. A model with $r_0\simeq \Delta r\simeq 0.2$ furthermore is consistent with the $1\sigma$ upper limit of BICEP, $r_{\rm BICEP}\simeq 0.27$, but in this case the tension with the full sky limit from Planck is not avoided. Still, this possibility could be constrained by future B-mode experiments.

While an explanation for the spatial variation of the tensor-to-scalar ratio points towards non-standard early-universe scenarios, the suggested model links two independent phenomena, providing a simple way to exclude this hypothesis.
It is furthermore clear that a simple dipolar scaling of the power modulation might not be sufficient \citep{Planck2013power}, or that even combinations of spatially varying cosmological parameters might be at work.
These aspects require more data and careful statistical tests which are beyond the scope of this paper. Measurements of the $EE$ and $BB$ power spectra as well as the $TE$, $TB$ and $EB$ power spectra would shed additional light on the underlying physical mechanism, allowing us to rule out different possibilities. In particular, the extra information could be used to increase the significance of a detection and push below the $TT$ cosmic variance limit if indeed $r$ is varying spatially at the level discussed here. Finally, even if spatially varying $r$ can only explain part of the hemispherical power asymmetry, it could render the scalar contribution to the anomaly less significant, dropping it below the $2\sigma$ level.

\vspace{-3mm}
\section{Conclusion}
\label{sec:conclusions}
We are entering a new era of CMB cosmology, with searches for B-modes turning into precise measurements. It is thus important to think about physical scenarios that can be tested with future polarization measurements, taking the clues given by the current data seriously. Here, we discussed the idea of a spatially varying tensor-to-scalar ratio connecting the recent BICEP2 result and the hemispherical power asymmetry. While this possibility requires a non-standard early-universe scenario, the model makes predictions that can be tested with future B-mode experiments which cover a large fraction of the sky. We argued that a spatial variation of $r$, while consistent with the current Planck full-sky limit as well as the BICEP2 result, cannot fully account for the amplitude of the hemispherical asymmetry. In fact, variations of $r$ as a full description of the asymmetry can be ruled out at more than $10\sigma$. 

However, a simple dipolar modulation of $r$ and even more complicated spatial dependencies are consistent with current measurements, and could still account for some portion of the hemispherical asymmetry. If it is present, the contribution from tensor fluctuations to the CMB polarization signal should be much smaller in the Northern hemisphere, pushing it close to $r\simeq 0$ in the extreme case, while suggesting a value for $r$ that is slightly larger than for BICEP2 towards the direction of the hemispherical power asymmetry maximum, $(l, b) \approx (227, -27)^\circ$. This would inevitably point us beyond the standard inflation scenario, providing a direct link for one of the CMB temperature anomalies with an underlying physical process, a possibility that should be further explored.

Future work will investigate the possible connection between this and other temperature anomalies, such as the quadrupole and octopole alignments \citep{Oliveira2004, Copi2004, Schwarz2004}. This is motivated by the fact that tensor modes also contribute to the temperature power spectrum at multipoles $\ell \lesssim 100$. It is furthermore important that spatial variations of tensor fluctuations are not constrained by large-scale structure surveys so that B-mode measurements can provide unique insights in this direction. Also, even if a spatial variation of $r$ can only explain part of the hemispherical power asymmetry, it could decrease the scalar contribution to the anomaly below the $2\sigma$ level.

\small

\vspace{-3.5mm}
\section*{Acknowledgments}
JC cordially thanks Glenn Starkman for stimulating discussions of this problem. He is also grateful to Saroj Adhikari, Grigor Aslanyan, Guido D'Amico, Anupam Mazumdar, Arthur Kosowsky and Subodh Patil for helpful comments on the manuscript. This work was supported by NSF Grant No. 0244990 and by the John Templeton Foundation.
 
\bibliographystyle{mn2e}
\bibliography{Lit}

\end{document}